\documentclass[aps,pra]{revtex4}

\def\duzomniejsze{<\kern-.7mm<}
\def\duzowieksze{>\kern-.7mm>}

\def\textbf#1{{\bf #1}}
\def\beq{\begin{equation}}
\def\eeq{\end{equation}}
\def\be{\begin{equation}}
\def\ee{\end{equation}}
\def\ben{\begin{eqnarray}}
\def\een{\end{eqnarray}}
\def\beqa{\begin{eqnarray}}
\def\eeqa{\end{eqnarray}}
\def\eea{\end{array}}
\def\bea{\begin{array}}
\newcommand{\bei}{\begin{itemize}}
\newcommand{\eei}{\end{itemize}}
\newcommand{\bee}{\begin{enumerate}}
\newcommand{\eee}{\end{enumerate}}

\def\ccal{{\cal C}}
\def\disteve{{\cal D}_E}
\def\distbob{{\cal D}_B}
\def\tr{{\rm Tr}}

\def\>{\rangle}
\def\<{\langle}
\def\blacksquare{\vrule height 4pt width 3pt depth2pt}
\def\ic{I_{coh}}
\def\ot{\otimes}

\def\dt#1{{{\kern -.0mm\rm d}}#1\,}

\def\rhota{\tilde \rho_A}
\def\rhotb{\tilde \rho_B}
\def\rhote{\tilde \rho_E}
\def\rhotx{\tilde \rho_X}

\def\rhonb{\rho_B'}
\def\rhone{\rho_E'}
\def\rhona{\rho_A'}
\def\rhonx{\rho_X'}

\def\rhotypa{{\hat \rho}_A^{typ}\,{}}
\def\rhotypb{{\hat \rho}_B^{typ}\,{}}
\def\rhotype{{\hat \rho}_E^{typ}\,{}}
\def\rhotypx{{\hat \rho}_X^{typ}\,{}}

\def\rhotypan{{\rho}_{A}^{typ}\,{}}

\def\sigalpb{{\sigma_B^\alpha}'}
\def\sigalpe{{\sigma_E^\alpha}'}
\def\sae{\sigma_E^\alpha}
\def\sab{\sigma_B^\alpha}

\def\phitabe{\tilde \phi_{ABE}}
\def\psitabe{\tilde \psi_{ABE}}
\def\psit{{\tilde \psi}}

\def\rk{{\rm rk\,}}

\def\supp{{\rm supp\,}}

\def\ep{\epsilon}

\def\lav{\bigl\<}
\def\rav{\bigl\>}

\def\bed{\begin{definition}}
\def\eed{\end{definition}}
\def\bet{\begin{theorem}}
\def\eet{\end{theorem}}

\def\bel{\begin{lemma}}
\def\eel{\end{lemma}}

\def\bep{\begin{proposition}}
\def\eep{\end{proposition}}

\newtheorem{lemma}{Lemma}

\newtheorem{theorem}{Theorem}
\newtheorem{proposition}{Proposition}

\begin{document}

\title{Quantum coding theorem from privacy and distinguishability}

\author{Micha\l{} Horodecki}
 \affiliation{
   Institute of Theoretical Physics and Astrophysics, 
   University of Gd\'{a}nsk, Poland
   }
   
\author{Seth Lloyd} 
 \affiliation{
 Massachusetts Institute of Technology, Cambridge, MA 02139 USA
   }   
   
\author{Andreas Winter}
 \affiliation{
     Department of Mathematics,
    University of Bristol,
    University Walk, Bristol BS8 1TW, U.~K.
    }

\begin{abstract}
We prove direct quantum coding theorem for random quantum codes. The problem
is separated into two parts: proof of distinguishability of codewords 
by receiver, and of indistinguishability of codewords by environment (privacy). 
For a large class of codes, only privacy has to be checked. 
\end{abstract}

\maketitle
\section{Introduction} 
The quantum coding theorem for transmission of quantum information via noisy 
channel  is one of the fundamental achievement of  quantum information. 
First proof, not yet fully rigorous,  of  direct coding theorem was given in \cite{Lloyd-cap}. 
The converse theorem was rigorously proven in \cite{BarnumNS-converse1997} 
(see \cite{HHH-cap2000} in this context).
An attmept to rigourous proof 
of direct coding theorem was subsequently done by Shor, with his notes published on website \cite{shor-cap}.
Later, Devetak provided first complete proof of coding theorem, using so called 
random CSS codes. Both for practical as well as for fundamental reasons 
it is desirable to have proof of coding theorem for fully random codes,
such as Lloyd, or Shor ones. 
In this paper we present a new proof of coding theorem. 
The main advantage of the present approach, 
is that we divide problem of coding theorem into two separate problems. 
One is whether Bob can distinguish signals, the second is whether 
environmenet cannot distinguish them.
It is well known that those two conditions are crucial for sending 
quantum information reliably. The first one is connected with bit error,
and the second with phase error \cite{Lloyd-cap,shor-cap,Devetak2003-coding,DevetakWinter-hash}.
Actually, Devetak's coding scheme is a coherent version of cryptographic 
protocol consisting  of two stages: error correction, 
that ensures distinguishability by Bob and privacy amplification 
responsible for diminishing Eve's knowledge. 

However so far the two problems have not been separated, in the sense,
that the proof of coding theorem did not consist of 
two completely separate mathematical problems. 
 

More specifically, we show that if we have set of $N$ vectors 
which, after crossing the channel are distinguishable for Bob (receiver),
but not for Eve (controlling environment), then Alice (sender) and Bob can share 
maximally entangled state of rank $N$. The statement of quantum coding theroem,
is that capacity is determined by coherent information  $\ic$ \cite{Schumacher-Nielsen}.  
Thus the question whether for $n$ uses of the channel, the set of vectors (quantum code)
that satisfy the two conditions  can have $2^{n\ic}$ elements.

In classical case the coding theorem states that capacity is determined by Shannon mutual 
information $I_M$. Since only distinguishability is needed, one proceeds as follows. 
One fixes source, and then pick codewords at random from the probability distribution 
of the source.  Such a randomly chosen code can have size as large as $2^{nI_M}$ 
and still be distinguishable after passing  the channel.

In quantum case  two problems arise. Firstly, as we have mentioned, we need 
two things: not only distinguishability  by Bob, but also indistinguishability by 
Eve (one can call it  "privacy"). Secondly, 
for a fixed source, unlike in classical case, we have many different ensembles
which are compatible with the source. 
Thus "picking random code" is not uniquely defined, so that we have to choose 
an ensemble. The advantage of our approach,
is that it allows to see that various ensembles can do the job. 
First of all, it turns out that for arbitrary ensemble, if we pick at random  
$2^{n\ic}$ vectors, they will be distinguishable for Bob. This we obtain  adapting a   one-shot version of  HSW theorem
\cite{Devetak-one-shot}. 
Thus to know that a given code is good, it remains to check 
privacy. It is easy to see that this will no longer hold for arbitrary ensemble.
To assure privacy, the ensemble must be reach enough. It turns out that 
checking privacy for some ensembles is a fairly easy task. 
In this paper we will do this for ensemble used by Lloyd, and for ensemble  generated by Haar measure, deformed by source density 
matrix. Most likely,  it also works for Gaussian codes (also deformed by source).

\section{Quantum error - analogue of classical error probability}
Here we will introduce a quantum parameter analogous to error probability in 
classical coding theorem. The main points have been already found 
in \cite{SchumacherW01-approx}, in a slightly different form. 

We need the following lemma, which says that when Alice is product 
with environment (Eve) then there exists Bob decoding, after which
he shares with Alice pure state which is purification of Alice's system
\begin{lemma}
For any pure tripartite state $\psi_{ABE}$, if reduced state $\rho_{AE}$ 
is product, then there exists unitary operation $U_{BB'}$, 
such that 
\be
I_{AE} \ot U_{BB'} |\psi_{ABE}\>|0\>_{B'}=|\psi_{AB}\> |\psi_{EB'}\>
\ee
\end{lemma}
{\bf Proof.} Follows from the fact that all purifications of a fixed state 
are related by a unitary transformation on (perhaps extended) ancilla. Here we take 
the state to be $\sigma_{AE}$ and extended ancilla system is $BB'$
Without loss of generality we have assumed here that the system $B$ is not smaller than $A$. 
\blacksquare

Here is version of the above lemma in approximate case.
\begin{lemma}
Consider a state $\psi_{ABE}$ and suppose that 
\be
||\sigma_{AE}-\sigma_A\ot \sigma_E||\leq \epsilon
\label{eq:qerror}
\ee
Then there exists unitary $U_{BB'}$ such that 
\be
F(\sigma'_{AB},\phi_{AB})\geq 1-{1\over 2}\ep
\ee
where $\sigma'_{AB}$ is reduced density matrix 
of state $I_{AE}\ot U_{BB'}|\psi_{AEB}\>|0\>_{B'}$ and 
$\phi_{AB}$ is purification of $\sigma_A$ (reduced density matrix of $\psi_{ABE}$); $||\cdot||$ denotes trace norm.
\label{lem:approx-prod}
\end{lemma}

{\bf Proof.} 
Using inequality $F(\rho,\sigma)\geq 1-{1\over 2} ||\rho-\sigma||$ \cite{Fuchs-Graaf}
we get  $F(\sigma_{AE},\sigma_A\ot \sigma_B)\geq 1 - \ep/2$. Then,  by definition 
of $F$, there exists purification $\phi_{ABB' E}$ such that 
\be
F(|\psi_{ABE}\>|0\>_{B'}, \phi_{A BB'E}) = F(\sigma_{AE},\sigma_A\ot \sigma_B)\geq 1 - {1\over 2}\ep
\ee
From the proof of previous lemma, we see that there exists unitary operation $U_{BB'}$ 
which factorizes state $\phi$ into $B$ and $B'$ (again we assume here that dimension of 
the system $B$ is no smaller 
than that of $A$). Thus 
\be
F(\psi'_{ABB'E},\phi'_{AB}\ot \phi'_{EB'})\geq 1-{1\over 2}\ep
\ee
where $|\psi'_{ABB'E}\>=I_{AE} \ot U_{BB'}|\psi_{ABE}\>\ot |0\>_{B'}$ 
and $\phi'_{AB}\ot \phi'_{EB'}= I_{AE} \ot U_{BB'}|\phi_{ABB'E}\>$.
Note that $\phi'_{AB}$ is purification of $\sigma_A$.
From monotonicity of $F$ under partial trace we get:
\be
F(\sigma'_{AB}, \phi_{AB}) \geq 1-{1\over 2} \ep
\ee
This ends the proof of the lemma. \blacksquare

It says that if Alice and Eve are approximately product, 
then there exists Bob's decoding, which restores with high fidelity 
of entanglement with Alice. 

Thus the parameter $||\sigma_{AE}-\sigma_{A} \ot \sigma_E||_{\tr}$ 
is what we could call "quantum error", an analogue of error 
probability in classical coding theorems. 

It is however convenient to consider modified quantum error,
which, if small, implies that  Alice and Bob share a state close 
to maximally entangled state of Schmidt rank  $N$. Henceforth in paper we will 
us this type of quantum error. It is given by 
\be
q_e=\|\sigma_{AE} - \tau_A \ot \sigma_E \| 
\label{eq:mod-qe}
\ee
where $\tau$ is normalized projector of rank $N$.  
From the above lemma we obtain the main result of this section:

\bep Consider arbitrary pure state $\psi_{ABE}$. Let $\sigma_{AB}$ 
be reduced density matrix of $\psi_{ABE}$. Then, 
for arbitrary $\ep>0$ if the quantum error (\ref{eq:mod-qe}) 
satisfies 
\be
q_e\leq \ep
\ee
then there exists Bob's operation $\Lambda_B$ such that 
\be
F(\sigma'_{AB},\psi^+_{AB})\geq 1-{1\over 2} \ep
\ee
where $\sigma'_{AB}=(I_A\ot \Lambda_B)\sigma_{AB}$, 
and  $\psi^+_{AB}={1\over \sqrt N}\sum_{i=1}^N |ii\>$.
\eep

Therefore, the task is to find a bipartite state for Alice, such that 
if she will send half of it down the channel, 
then the quantum error will be small, for $N\simeq 2^{n \ic}$. 
We will construct such a state from a code. Given a code $\{|\alpha\>\}_{\alpha=1}^N$ 
the state will be 
\be
\psi_{AA'}={1\over \sqrt N} \sum_{\alpha=1}^N|\psi_\alpha\>|\alpha\>
\ee
with $\{\psi_\alpha\}$ being orthonormal set of vectors.
The task is thus to show that, if $A'$ is sent down the channel,
then $q_e$ is small. 

\section{Distinguishability and privacy imply small quantum error}

The channel from Alice to Bob implies dual channel to Eve, who represents environment.
Thus if Alice sends a state $|\alpha\>$,
we obtain two kinds of output states: Bob's output state $\sigma^\alpha_B$ 
and Eve's output state $\sigma^\alpha_E$. In this section we will 
show that a set of states $\ccal=\{|\alpha\>\}$ for which Bob's output 
states are approximately distinguishable, while Eve's output 
states are not, then the set $\ccal$ is a good quantum code. 
This means that if Alice will create state 
\be
\psi_{AA'}={1\over \sqrt N} \sum_{\alpha=1}^N|\psi_\alpha\>|\alpha\>
\ee
where $\psi_\alpha$ are orthonormal vectors. 
and send the system $A'$ down the channel,
then the quantum error for arising tripartite state $\psi_{ABE}$ 
will be approximately zero. The distinguishability we will quantify by Holevo function.

Before we formulate suitable proposition, let us express output states 
of Bob and Eve in terms of the joint state $\psi_{ABE}$ obtained as described above. 
If we perform measurement onto basis  $\psi_\alpha$,
the state of the systems $BE$ will collapse to the state $\psi^\alpha_{BE}$. 
The states $\sigma^\alpha_B$ and $\sigma^\alpha_E$ are then given by 
\be
\sigma^\alpha_B=\tr_E (|\psi_\alpha\>\<\psi_\alpha|_{BE}),\quad 
\sigma^\alpha_E=\tr_B (|\psi_\alpha\>\<\psi_\alpha|_{BE})
\ee

Let us now define more precisely distinguishability and privacy.
For a set of $N$ states $\ccal=\{|\alpha\>\}$ and channel $\Lambda$ 
let us define  distinguishability by Eve $\disteve$ and distinguishability $\distbob$ as follows 
\ben
&&\disteve= \chi_E \\ 
&&\distbob= \chi_B 
\label{eq:priv-dist}
\een
where $\sigma_{E,B}^\alpha$ are Eve's and Bob's output states 
defined above and  
$\chi_E= \chi(\{{1\over N},\sigma_E^\alpha\}), 
\chi_B= \chi(\{{1\over N},\sigma_B^\alpha\})$.
Here $\chi(\{p_i,\rho_i\})=S(\sum_ip_i\rho_i)-\sum_i p_i s(\rho_i)$
with  $S(\rho)=\tr \rho\log \rho$ being von Neumann entropy.

\begin{proposition}
The quantum error satisfies the following inequality 
\be
q_e\leq c \sqrt{\disteve + (\log N- \distbob)}
\ee
where $c=\sqrt{2\ln 2}$. 
\end{proposition}
{\bf Proof.} Let subsystems of $\psi_{ABE}$  are denoted by $\sigma_X$ 
and their entropies by $S_X$ with suitable subscript.
Because $\sigma_B^\alpha$ and $\sigma_E^\alpha$ 
are reduced density matrices of the same state $\psi_{BE}^\alpha$ 
we have 
\be
S(\sae)=S(\sab)
\ee
Thus 
\be
\chi_B -\chi_E = S_B -S_E
\ee
so that we obtain 
\be
\distbob + (\log N- \disteve) = \log N + S_E - S_{AE}=S(\sigma_{AE} | \tau_A \ot \sigma_E)
\ee
where we have used the fact that total state is pure so that $S_{AE}=S_B$,
and $\tau_A=I/N$ is maximally mixed state on the system $A$. 
Now, using well known relation $S(\rho|\sigma) \geq {1\over 2\ln 2} \|\rho-\sigma\|^2$ 
\cite{OhyaPetz} we obtain the required inequality. 
This ends the proof of the proposition.\blacksquare

{\bf Remark.} Essentially, we have merged two facts. 
First, as obtaind in \cite{SchumacherW01-approx} 
that when $I_{coh}$ is close to entropy of the source, 
then the error correction condition is approximately satisfied
(i.e. $q_e$ is small). Second, that 
$\chi_B-\chi_E=I_{coh}$  which was exploited in \cite{Devetak2003-coding,DevetakWinter-hash}.

\section{Overview}
In spirit of Shannon we will consider mental construction: a source $\rho_A^{\ot n}$,
and a joint state $\phi_{ABE}$
which emerges from sending half of a purification of $\rho_A^{\ot n}$ 
down the channel (as before, $E$ represents environment).
We will also use the state $\phi$ projected onto typical subspaces. 
This will be denoted by $\phi_{ABE}'$. The two states $\phi$ and $\phi'$ 
will be shown to be close to each other in trace norm. 

A code will be picked at random from ensemble 
which gives rise to typical version of $\rho_A$. 
Depending on chosen ensemble, we will have different types of codes. 
For any code $\{|\alpha\>\}$ we will consider a
bipartite state of Alice of the form $\psi_{AA'}= {1\over \sqrt N} 
\sum_{\alpha=1}^N |\psi_\alpha\> |\alpha\>$
with $\{\psi_\alpha\}$ being orthonormal set of vectors. 
This is the {\it actual} state 
that will be sent down the channel. 
The resulting joint state of Alice, Bob and Eve will be denoted by $\psi_{ABE}(\ccal)$
(shortly $\psi_{ABE}$). Again a version of this state, projected onto typical subspaces 
will be denoted by $\tilde\psi_{ABE}$. We will show that for certain codes 
(i.e. for certain ensembles) the states $\psi$ and $\tilde\psi$ are close 
to each other in trace norm. 

Since the states are close to each other, it is enough to consider 
privacy and distinguishability for the projected state. 
Interestingly, we will show that distinguishability is then merely 
a consequence of the  fact that the ensemble from which we choose the code
gives rise to typical version of $\rho_A$. This we obtain by 
adapting one-shot version of HSW thereom proven by Devetak \cite{Devetak-one-shot}. 
Privacy should be checked case by case for different codes. 

\section{Source, channel  and typicality}

\label{sec:source}
Let us fix a source $\rho_A^{\ot n}$. Half of purification of the source 
is sent down the channel 
\be
\Lambda^{\ot n}(\cdot)=\sum_{k} A_k (\cdot) A_k^\dagger
\ee
(it is not actual sending but a mental construction, as in classical coding theorem).
This creates pure state $\phi^{\ot n}_{ABE}$ shared by Alice,
Bob and environment.  Explicitly  we have 
\be
\phi_{ABE}= \sum_{i\in I}\sum_{k\in K}\sqrt{p_i} |i\>_A A_k |i\>_B|k\>_E
\ee
where $\rho_A^{\ot n}=\sum_{i\in I} p_i |i\>\<i|$, 
$I$ is set of indices of complete eigenbasis of $\rho_A^{\ot n}$, 
and $K$ is set of indices of complete eigenbasis of $\rho_E^{\ot n}$.
Later we will consider typical subspaces, and this we will indicate by 
omitting $I$ and $K$. 

Subsystems of this state are $\rho_A^{\ot n}$,
$\rho_B^{\ot n}=\Lambda^{\ot n}(\rho_A^{\ot n})$ and $\rho_E^{\ot n}$.
Projectors onto typical subspaces of these states are respectively:
$\Pi_A$, $\Pi_B$ and $\Pi_E$. The (unnormalized) typical versions of the states 
we denote by $\rhotypa=\Pi_A \rho_A^{\ot n} \Pi_A$, $\rhotypb=\Pi_B \rho_B^{\ot n} \Pi_B$,
$\rhotype=\Pi_E \rho_E^{\ot n} \Pi_E$. Explicitly we have 
\ben
&&\rhotypa=\sum_i p_i |i\>\<i| \\ 
&&\rhotypb=\Pi_B\sum_{k\in K} A_k \rho_A^{\ot n} A_k^\dagger \Pi_B \\ 
&&\rhotype=\sum_{k,k'} \tr (A_k  \rho_A^{\ot n} A_{k'}^\dagger)|k\>\<k'|
\een
In second equation  we write $k\in K$ because  the sum runs over all 
vectors $|k\>$, while in first and third equation the sum runs only over 
the set of indices corresponding to  vectors from typical subspace. 
We have to explain why typical projector $\Pi_E$ 
is given by $\sum_k |k\>\<k|$. This follows form the fact that the 
Kraus operators can be chosen in such a way that 
the state $\rho_E^{\ot n}$ is diagonal in basis $\{|k\>\}$.  We assume that our Kraus operators 
are such ones (so that the only diagonal terms of the last equation are 
in fact nonzero).

We have lemma \cite{Schumacher1995,svw2005,CoverThomas}

\begin{lemma} (typical states)
\label{lem:typical}
For arbitrary $0<\ep<1/2$, $\delta>0$, for all $n$ large enough we have 
that for $X=A$, $X=B$ and $X=E$: 
\bee
\item The states $\rhotypx$ are almost normalized i.e. 
\be
\tr \rhotypx\geq 1-\ep
\label{eq:lem-typ-tr}
\ee
\item All eigenvalues $\lambda$ of $\rhotypx$ satisfy
\be
2^{-n(S_X+\delta)}\leq \lambda\leq 2^{-n(S_X-\delta)}
\label{eq:lem-typ-eig}
\ee
\item The ranks of states $\rhotypx$ satisfy
\be
\rk (\rhotypx) \leq 2^{n(S_X +\delta)}
\label{eq:lem-typ-rk}
\ee
\eee
\end{lemma}

{\bf Remark.} Both in this lemma, and in  Prop. \ref{prop:rhonx}, 
$\ep$ can be exponential in $n$, i.e. the relations hold with 
\be
\ep=e^{-c\delta^2 n}
\ee
where $c$ is a constant. 

Let us now consider an unnormalized state 
\be
|\phitabe\>=\Pi_A \ot \Pi_B \ot \Pi_E|\phi_{ABE}\>
\ee
The state can be written
\be
|\phitabe\>=\sum_{i,k} \sqrt{p_i}|i\>_A F_k|i\>_B |k\>_E
\ee
where $F_k=\Pi_B A_k$.  

The reductions of this state are a bit different than typical states $\rhotypx$. 
We will denote them by $\rhota$, $\rhotb$ and $\rhotb$. 
We have 
\ben 
&&\rhota= \sum_ip_i |i\>\<i|\tr(\sum_k F_k|i\>\<i|F_k^\dagger)\\
&&\rhotb=\Pi_B\sum_k A_k \rhotypa A_k^\dagger \Pi_B=\sum_k F_k \rhotypa F_k^\dagger\\
&&\rhote=\sum_{k,k'} \tr(F_k \rhotypa F_{k'}^\dagger) |k\>\<k'|
\een

Finally, we consider {\it normalized} typical Alice's state $\rhotypan$
given by 
\be
\rhotypan = {1\over \tr \rhotypa} \rhotypa
\ee
We then modify the above states $\rhota$, $\rhotb$ and $\rhote$ 
into still unnormalized states 
\ben 
&&\rhonb={1\over \tr \rhotypa}\rhotb=\Pi_B\sum_k A_k \rhotypan A_k^\dagger \Pi_B=\sum_k F_k \rhotypa F_k^\dagger\nonumber\\
&&\rhone={1\over \tr \rhotypa}\rhote=\sum_{k,k'} \tr(F_k \rhotypan F_{k'}^\dagger) |k\>\<k'|\nonumber\\
&&\rhona= {1\over \tr \rhotypa} \sum_ip_i |i\>\<i|\tr(\sum_k F_k|i\>\<i|F_k^\dagger)
\een

\subsection{Properties of states $\rhonx$}
\label{subsec:rhonx}

The states that we will use most frequently in the proof are $\rhonx$.
We have the following proposition
\begin{proposition}
\label{prop:rhonx}
For arbitrary $0<\ep<1/2$, $\delta>0$, for all $n$ large enough we have 
that for $X=A$, $X=B$ and $X=E$: 
\bee
\item The states $\rhonx$ are almost normalized i.e. 
\be
\tr \rhonx\geq 1-\ep.
\label{eq:prop-rhonx-tr}
\ee
\item  The eigenvalues $\lambda$ of $\rhonx$ satisfy
\be
\lambda\leq (1+\ep) 2^{-n(S_X-\delta)}
\label{eq:prop-rhonx-eig}
\ee
\item The ranks of states $\rhonx$ satisfy
\be
\rk (\rhonx)\leq 2^{n(S_X +\delta)}.
\label{eq:prop-rhonx-rk}
\ee
\item The following inequality holds
\be
\tr \rhonx^2\leq (1+\ep) 2^{(-n{S_X-\delta})}.
\label{eq:prop-rhonx-tr2}
\ee
\eee
\end{proposition}

{\bf Proof.} 
We first note that $\rhotx\leq \rhotypx$, which holds, 
because $\rhotx$ can be obtained from $\rhotypx$ by projecting onto 
second subsystem of bipartite state, whose reduction is $\rhotx$. 
Using lemma \ref{lem:typical}, eq. (\ref{eq:lem-typ-tr}) we then obtain that for large $n$
\be
\rhonx\leq (1+{\ep\over 4}) \rhotypx
\ee
This via lemma \ref{lem:typical} immediately gives eq. (\ref{eq:prop-rhonx-tr}) and 
eq. (\ref{eq:prop-rhonx-eig}) as well as (\ref{eq:prop-rhonx-rk}). 
To prove (\ref{eq:prop-rhonx-tr2}) we note that $0\leq X\leq Y$ implies
$\tr X^2  \leq \tr Y^2$. \blacksquare

There follow useful expressions \cite{Lloyd-cap}
\ben
&&\sum_{k,k'} \tr (F_k \rhotypan F_{k'}^\dagger)\tr (F_{k'} 
\rhotypan F_{k}^\dagger) =\tr \rhone^2\\
&&\sum_{k,k'} \tr (F_k \rhotypan F_{k}^\dagger F_{k'} \rhotypan F_{k'}^\dagger) =\tr \rhonb^2
\een
One also defines matrix 
\be
\rho_{i/o}= {1\over \tr \rhotypa} \sum_{i,k} p_i |i \>_A\< i| \ot F_k( |i\>_B\<i|) F_k^\dagger
\ee
One finds that reduced density matrices of $\rho_{i/o}$ 
are $\rhona$ and $\rhonb$. One can also find by direct checking 
that 
\be
\tr \rho_{i/o}^2 \leq \min( \tr \rhonb^2,\tr \rhona^2).
\label{eq:rhoio-rhonx}
\ee
\section{Random codes}
\label{sec:random-codes}

As we have mentioned, in quantum case there is no unique way of drawing codes at 
random.  In this paper we will use two types of quantum random codes: Lloyd codes  defined 
in \cite{Lloyd-cap} and uniform, source-distorted codes 
For those codes 
we will show that the average quantum error is small, if we choose $2^{nR}$ codewords,
with $R<I_{coh}$.  

{\it Lloyd codes}: The codes of \cite{Lloyd-cap} are defined as follows. 
To have a code consisting of $N$ vectors, 
Alice picks  $N$ vectors according the following distribution
\be
|\alpha\>=\sum_i \sqrt{q_i} e^{i \phi_i} |i\>
\label{eq:alpha}
\ee
where $q_i$ and $|i\>$ are eigenvalues and eigenvectors  of $\rhotypan$,
i.e. $q_i=p_i/\tr \rhotypa$.
The phases $\phi_i$ are drawn independently and uniformly from unit circle.
Average over such $\alpha$'s we will denote by $\int (...) \dt \alpha$.
Note that the ensemble defining Lloyd codes give rise to typical version of $\rho_A^{\ot n}$
\be
\int |\alpha\>\<\alpha| \dt \alpha = \rhotypan.
\ee
{\it Uniform source-distorted codes}. Alice picks 
$N$ vectors according to the following distribution 
\be
|\alpha\>=\sqrt{d_A} \sqrt {\rhotypan} |\phi\>
\ee
where $|\phi\>$ is taken uniformly from typical subspace of system $A$ 
(i.e. give by projection $\Pi_A$. The codewords are not of unit length
but with high probability they are almost normalized.
We prove it by Chebyshev inequality. Compute 
variance of $\<\alpha|\alpha\> = d_A\<\phi|\rhotypan|\phi\>$ obtaining 
\be
Var={d_A^2 \over d_A^2 +d_A} \tr (\rhotypan)^2 +
{d_A^2 \over d_A^2 +d_A} \tr \rhotypan - 1  
\leq \tr (\rhotypan)^2 \leq (1+\ep) 2^{-n (S_A -\delta)}
\label{eq:var-andr}
\ee
which gives  
\be
Prob(\<\alpha|\alpha\> \not \in (1-\ep,1+\ep)) \geq 1-{(1+\ep)\over \ep^2} 2^{-n (S_A -\delta)}
\label{eq:Acodes-norm}
\ee
The last inequality comes from lemma \ref{lem:typical}, and holds for all $n$ large enough.
If we take $N=2^{nR}$ with $R<\ic$, we get that for arbitrarily fixed $\ep$ 
the probability  of failure goes exponentially down for all codewords. 
However we will not use so strong result. It will be enough 
to know that a randomly picked codeword with high probability  has norm close to 
$1$.  Finally, note that, again we have  
\be
\int |\alpha\>\<\alpha| \dt \alpha = \rhotypan.
\ee

\section{Joint state $\psi_{ABE}$ for a fixed code sent down that channel.} 
\label{sec:psi-actual}

In this section we discuss properties of the actual state that can be obtained
by use of  codes. The main result  of this section is that 
with high probability, the state $\psi_{ABE}$ is close to 
its version projected onto typical subspcaces. 
We will separately discuss the case of normalized and unnormalized codes.
For normalized codes, the result follows solely from 
the fact the ensemble of the code  gives rise to $\rhotypan$. 
For unnormalized codes one has to check it case by case (we will check it here for 
uniform source-distorted codes). 

\subsection{Normalized codes}
Having fixed a code $\ccal=\{|\alpha\>\}_{\alpha=1}^N$,
Alice creates state 
\be
\psi_{AA'}={1\over \sqrt N} \sum_{\alpha=1}^N |\psi_\alpha\>_A\ot|\alpha\>_{A'}
\label{eq:psiaa}
\ee
where $\{|\psi_\alpha\>\}_{\alpha=1}^N$ is orthonormal set of $N$ vectors. 
Then she sends $A'$ down the channel $\Lambda^{\ot n}$ to Bob. 
The emerging state $\psi_{ABE}$  is the following:
\be
\psi_{ABE}=\psi_{ABE}(\ccal) ={1\over \sqrt N } \sum_{\alpha}\sum_{k\in K} |\psi_\alpha\>_A 
A_k|\alpha\>_B |k\>_E
\label{eq:psiabe}
\ee
Subsequently, we consider projected version of the state  
\be
|\psitabe\>= I_A \ot \Pi_E\ot \Pi_B |\psi_{ABE}\>.
\ee
This state can be written as 
\be
|\psitabe\>={1\over \sqrt N } \sum_{\alpha,k} |\psi_\alpha\>_A 
F_k|\alpha\>_B |k\>_E
\label{eq:psitabe}
\ee
where recall that $F_k=\Pi_B A_k$, and we sum only over typical indices $k$.

Let us now prove that with high probability, the state $\psitabe$ will be close to 
the actual state $\psi_{ABE}$ shared by Alice, Bob and Eve.

\begin{proposition}
\label{prop:typicality-nor}
For arbitrary $0<\ep<1$, for $n$ large enough we have 
\be
||\, |\psi_{ABE} \>\<\psi_{ABE}| - |\psitabe\>\<\psitabe|\,||\leq \ep
\ee
with high probability for any normalized codes satisfying $\int |\alpha\>\<\alpha|\dt alpha
=\rhotypan$
\end{proposition}
{\bf Proof.}
From lemma \ref{lem:psi-psit} it follows that it is enough to prove 
that for $n$ large enough we have 
\be
|\<\psitabe|\psi_{ABE}\>|^2\geq 1-\ep
\ee
We will show now that it is true with high probability. 
We have 
\be
|\<\psi|I_E \ot \Pi_B \ot \Pi_E |\psi\>|^2={1\over N} \sum_\alpha 
\tr( \sum_k F_k^\dagger F_k |\alpha\>\<\alpha|) 
\ee
We note that 
\be
\lav |\<\psi|I_E \ot \Pi_B \ot \Pi_E |\psi\>|^2 \rav_\alpha=
\tr \sum_k F_k \rhotypan F_k^\dagger=\tr \rhonb
\ee
From proposition \ref{prop:rhonx}, Eq. (\ref{eq:prop-rhonx-tr}) it follows that for 
$n$ large enough we have  $\tr \rhonb \geq 1-\ep^2$. 
The random variable $1- |\<\psi|I_E \ot \Pi_B \ot \Pi_E |\psi\>|^2$ 
is nonnegative, so we can use Markov inequality obtaining
\be
Prob( |\<\psi|I_E \ot \Pi_B \ot \Pi_E |\psi\>|^2\leq 1-2\ep)\leq  \ep
\ee
This ends the proof. \blacksquare

{\bf Remark.} Note that in proof we have only used the fact that 
codewords are normalized, and picked from an ensmeble whose 
density matrix is $\rhotypan$. 

\subsection{Unnormalized codes}
If codewords are not normalized, 
Alice prepares the following state:
\be
\psi_{AA'}={1\over \sqrt{ \sum_{\alpha=1}^N\<\alpha|\alpha\>}} 
\sum_{\alpha=1}^N |\psi_\alpha\>_A\ot|\alpha\>_{A'}
\label{eq:psiaa2}
\ee
The emerging Alice, Bob and Eve state is then 
\be
\psi_{ABE} ={1\over  
\sqrt{ \sum_{\alpha=1}^N\<\alpha|\alpha\>} } \sum_{\alpha}\sum_{k\in K} |\psi_\alpha\>_A 
A_k|\alpha\>_B |k\>_E
\label{eq:psiabe2}
\ee

We will consider unnormalized version of this state
\be
\psi'_{ABE} ={1\over \sqrt N } \sum_{\alpha}\sum_{k\in K} |\psi_\alpha\>_A 
A_k|\alpha\>_B |k\>_E
\label{eq:psiabe3}
\ee
and project it onto typical subspaces, obtaining 
\be
|\psitabe\>={1\over \sqrt N } \sum_{\alpha,k} |\psi_\alpha\>_A 
F_k|\alpha\>_B |k\>_E.
\label{eq:psitabe2}
\ee
We will now show that for uniform source-distorted  codes the state $\psitabe$
is close in trace norm   to the actual state  
$\psi_{ABE}$ shared by Alice and Bob.


\begin{proposition}
\label{prop:typicality-unnor}
For arbitrary $0<\ep<1$, for $n$ large enough we have 
\be
||\, |\psi_{ABE} \>\<\psi_{ABE}| - |\psitabe\>\<\psitabe|\,||\leq \ep
\ee
with high probability for uniform source-distorted codes.
\end{proposition}

{\bf Proof.} From lemma \ref{lem:psi-psit} it follows 
that we have to show that with high probability we have 
\be
|\<\psi_{ABE}'|\psitabe\>|^2\geq 1-\ep 
\label{eq:psiprim-psit}
\ee
and 
\be
\<\psi_{ABE}'|\psi_{ABE}'\> \leq 1+\ep.
\label{eq:psiprim}
\ee
The last inequality reads as 
\be
{1\over N} \sum_\alpha \<\alpha|\alpha\> \leq 1+\ep
\ee
We will now compute average and variance of this quantity over codes. 
Using (\ref{eq:var-andr}) we get 
\be
\lav {1\over N} \sum_\alpha \<\alpha|\alpha\> \rav_\ccal =\lav \<\alpha|\alpha\> \rav_\ccal =1
\ee
and 
\be
Var=Var(\<\alpha|\alpha\>)\leq (1+\ep)  2^{-n(S_A -\delta)}
\ee
since codewords are picked independently. 
 Thus form Chebyshev inequality we get 
\be
Prob(|{1\over N}\sum_\alpha \<\alpha|\alpha\> - 1| \geq \ep)\leq  {1+\ep \over \ep^2}\ 2^{-n(S_A -\delta)}
\ee
Thus with high  probability the inequality (\ref{eq:psiprim}) is satisfied.
To prove the same for inequality (\ref{eq:psiprim-psit}) 
we write 
\be
|\<\psi'|I_E \ot \Pi_B \ot \Pi_E |\psi'\>|^2={1\over N} \sum_\alpha 
\tr( \sum_k F_k^\dagger F_k |\alpha\>\<\alpha|)\equiv {1\over N} \sum_{i=1}^N X_i 
\ee
where $X_i$ defined by the above equality are i.i.d random variables. 
We note that 
\be
\lav X_i \rav_\alpha=\tr \sum_k F_k \rhotypan F_k^\dagger=\tr \rhonb\geq 1-\ep
\ee
We compute variance. 
\be
Var({1\over N} \sum_i X_i) = Var X_i=
\lav \<\alpha|Y |\alpha\> \<\alpha Y |\alpha\> \rav_\alpha 
-(\tr \rhonb)^2
\ee
where $Y=\sum_k F_k^\dagger  F_k$. 
For particular case of uniform source-distorted codes we obtain 
\be
Var\leq \tr Y\rhotypan Y\rhotypan + (\tr Y\rhotypan)^2 -(\tr \rhonb)^2= \tr \rhonb^2 \leq 
(1+\ep)2^{-n(S_B-\delta)}
\ee
where we have used $\tr (\sum_k F_k^\dagger F_k \rhotypan)=\tr \rhonb$.
Using Chebyshev inequality we get 
\be
Prob(|\<\psitabe|\psi_{ABE}\>|^2 \leq  1-\ep ) \leq {1-\ep\over \ep^2} \  2^{-n(S_B-\delta)}
\ee
This proves that with high probability, also the inequality (\ref{eq:psiprim-psit})
is satisfied.
This ends the proof. \blacksquare

\section{Coding theorem via distinguishability and privacy.}

In this section we prove the main result of this paper:
direct coding theorem by use of distinguishability and privacy. 
Our proof will show that for normalized random codes, 
only two conditions assure that the codes can be used in coding theorem:
\bei
\item[1)] The ensemble gives rise to $\rhotypan$. 
\item[2)] Privacy: codewords produce indistinguishable Eve's output states. 
\eei
We see that these  are quite modest conditions.

\subsection{Bob's and Eve's output states.}
It is useful to write down Alice and Bob's output states in terms 
of Kraus operators. We have 
\ben
&&\sigma_\alpha^E=\sum_{k,k'\in K} \tr (A_k |\alpha\>\<\alpha| A_{k'}^\dagger) |k\>\<k'| \\
&&\sigma_\alpha^B=\sum_{k\in K}  (A_k |\alpha\>\<\alpha| 
A_k^\dagger) \equiv \Lambda^{\ot n}(|\alpha\>\<\alpha|) \\
\een
Now, we introduce modified 
states  $\sigalpe$ and $\sigalpb$ given by 
\ben
&&\sigalpe=\sum_{k,k'} \tr (F_k |\alpha\>\<\alpha| F_{k'}^\dagger) |k\>\<k'| \\
&&\sigalpb=\sum_{k}  F_k |\alpha\>\<\alpha| 
F_k^\dagger 
\een
These states are reductions of the state (\ref{eq:psitabe}) and  (\ref{eq:psitabe2}), respectively. 
For normalized codes, the states are subnormalized, 
while for unnormalized codes, their trace may be both below and above 1. 
They have the following properties. 
For any code satisfying $\int |\alpha\>\<\alpha| \dt \alpha = \rhotypan$ 
we obtain 
\be
\int \sigalpe\dt \alpha=\rhone,\quad \int \sigalpb\dt \alpha=\rhonb
\ee
Moreover due to propositions \ref{prop:typicality-nor} and \ref{prop:typicality-unnor},  
and monotonicity of trace norm under trace preserving CP maps,
they are close on average to the original 
states $\sigma_\alpha^E$, $\sigma_\alpha^B$, which is stated in the following lemma 
\begin{lemma}
\label{lem:sig-sig}
For arbitrary $0<\ep<1/2$ and for $n$ high enough, 
with high probability we have 
\be
{1\over N} \sum_\alpha ||\sigma_\alpha^E - \sigalpe||\leq \ep, \quad   
{1\over N} \sum_\alpha ||\sigma_\alpha^B - \sigalpb||\leq \ep.
\label{eq:sig-sig}
\ee
where $\ep$ is independent on $N$, and can be taken to be exponential in $n$.
\end{lemma}

Recall, that $\ep$ is exponential in $n$. 

\subsection{Privacy}
We first prove that if $N=2^{nR}$ with $R<\ic$ 
then the distinguishability by Eve is arbitrarily small. 
Let us first  bound  $\chi_E$ by average norm distance. 
We apply Fannes inequality \cite{Fannes1973}  
\be
|S(\rho) -S(\sigma)|\leq  ||\sigma-\rho||\log d + \eta(||\sigma -\rho||)
\ee
where $\rho$ and $\sigma$ are any states satisfying $||\rho-\sigma||<1/3$, $\eta(x)=-x \log x $,
and $d$ is dimension of the Hilbert space.
Denoting ${1\over N}\sum_{\alpha\in\ccal}||\sigma_E^\alpha -\sigma_E||=x$,
and using convexity of $\eta$  we obtain 
\be
\chi_E =\chi_E\leq x n \log d  + \eta(x)
\ee
where $d$ is input dimension of Hilbert space of $\rho_A$ (i.e. it is a constant).
Thus if we show that ${1\over N}\sum_{\alpha\in\ccal}||\sigma_E^\alpha -\sigma_E||$ 
is exponentially small for rate $R<\ic$, 
then also $\chi_E$ will be exponentially small.

\begin{proposition}(privacy). For the state $\psi_{ABE}$ and $N=2^{nR}$ 
where $R<\ic$ with high probability 
(over codes) we have 
\be
{1\over N}\sum_{\alpha\in\ccal}||\sigma_E^\alpha -\sigma_E||\leq \epsilon.
\ee 
\end{proposition}
{\bf Proof.} First we note that 
\be
||\sigma_E^\alpha -\sigma_E||\leq 
||\sigma_E^\alpha -\rhone|| + ||\sigma_E -\rhone||
\leq 2||\sigma_E^\alpha -\rhone|| 
\ee
In the last inequality we have used convexity of norm, and the fact that $\sigma_E$ 
is mixture of $\sigma_E^\alpha$. Subsequently we have 
\be
||\sigma_E^\alpha -\rhone||\leq ||\sigma_E^\alpha - \sigalpe|| + ||\sigalpe -\rhone||
\ee
Using (\ref{eq:sig-sig}) we finally get 
\be
{1\over N} \sum_\alpha ||\sigma_E^\alpha - \sigma_E||\leq 2\ep + 
{1\over N} \sum_\alpha ||\sigalpe-\rhone||
\label{eq:sig-sig-sum}
\ee
Thus we have to estimate $||\sigalpe -\rhone||$.
To this end we use lemma \ref{lem:norms} and get 
\be
||\sigalpe -\rhone||^2\leq d \tr (\sigalpe -\rhone)^2
\label{eq:norms-e}
\ee
where $d$ is dimension of the Hilbert space on which both $\rhone$ 
and $\sigalpe$ act. We note that 
\be
\sigalpe =\Gamma(|\alpha\>\<\alpha|),\quad \rhone=\Gamma(\rhotypan)
\ee
where $\Gamma$ is a CP map.
Then due to lemma \ref{lem:supp} we have 

Thus we have 
\be
\supp(\sigma_\alpha)\subset\supp(\rhone)
\ee
so that the dimension $d$ can be chosen as dimension of support of $\rhone$.
Thus due to Eq. (\ref{eq:prop-rhonx-rk}) which says that rank of 
$\rhone$ is bounded by rank of $\rhotype$ we have 
\be
d\leq 2^{n(S_E+\delta)}
\label{eq:de}
\ee
Now, we compute average square of Hilbert-Schmidt distance for Seth's codes
\be
\lav \tr (\sigalpe -\rhone)^2\rav_\alpha = \lav \tr (\sigalpe)^2 \rav_\alpha
+\tr \rhone^2 -2 \lav \tr(\sigalpe\rhone)\rav_\alpha 
\ee
We have 
\be
\lav \tr (\sigalpe)^2\rav_\alpha= \sum_{kk'}\int 
\<\alpha|F_{k'}^\dagger F_k|\alpha\>\<\alpha|F_k^\dagger F_{k'}|\alpha\>
=\tr \rhone^2 + \tr \rhonb^2 - \tr \rho_{i/o}^2\leq \tr \rhone^2 + \tr \rhonb^2 
\ee
and 
\be
\lav \tr \sigalpe\rhone\rav_\alpha = \tr \rhone^2
\ee
So that 
\be
\lav \tr (\sigalpe -\rhone)^2\rav_\alpha \leq \tr \rhonb^2 \leq 2^{-n(S_B-\delta)}
\label{eq:hs-sb}
\ee
The same we obtain for uniform source-distorted codes.  Then 
from eqs. (\ref{eq:de}), (\ref{eq:hs-sb})  and (\ref{eq:norms-e})  we get
\be
\lav ||\sigalpe - \rhone||^2\rav_\alpha\leq 2^{-n(\ic+\delta)}.
\ee
By Markov inequality we then obtain 
\be
Prob(||\rho_E^\alpha - \rhone||^2\geq \ep^2)\leq {1\over \ep^4 } 2^{-n(\ic+\delta)}
\ee
Due to equation (\ref{eq:sig-sig-sum})  we get 
\be
{1\over N} \sum_\alpha ||\sigma_E^\alpha - \sigma_E||\leq 3\ep
\ee
with high probability over codes. 
This ends the proof of the proposition. Recall, that $\ep$ can 
be taken exponential in $n$. \blacksquare

\subsection{Distinguishability}
Here we prove that Bob's outputs are distinguishable for $R<\ic$. 


To show that $\chi_B$ is close to $\log N$ 
it is enough that for a random code with $R<\ic$ is distinguishable by some  measurement,
with probability of error exponentially small. 
This can be shown again by referring to Fannes inequality.
Let outcomes of POVM be denoted by $\beta$. 
Then by Fannes inequality we obtain that the mutual information 
between input signals $\alpha$, and the measurement output 
satisfies 
\be
I(\alpha:\beta)\geq \log N - (8 p_e n \log d + 3 \eta(p_e))
\ee
where $p_e$ is probability of error
\be
p_e=1-{1\over N} \sum_{\alpha=\beta} p(\beta|\alpha)
\ee
with $p(\beta|\alpha)$ is probability of obtaining outcome $\beta$ given the 
state was $\sigalpb$. 
Since postprocessing can only decrease mutual information, 
we have 
\be
\chi_B' \geq I(\alpha:\beta).
\ee
Thus, if we can show that for a random code, Bob's output states are 
distinguishable by some measurement with exponentially small (in $n$)
probability of error.

To this end we will modify a result due to \cite{Devetak-one-shot}:
\begin{lemma} Assume that an ensemble $\{p_i,\rho_i:i\in S\}$ with $\sum_ip_i\rho_i=\rho$
such that there exist projectors $\Pi_i$, $\Pi$ satisfying
\ben
\tr \rho_i P \geq 1-\ep \\
\tr \rho_i \Pi_i \geq 1-\ep \\
\tr \Pi_i \leq 2^{nL} \\
\Pi \rho \Pi \leq 2^{-nG} \Pi
\een
Then there exists a subset $\ccal$ of  $S$ of size $2^{n[G -L - \delta]}$ 
and a corresponding POVM  $\{ Y_i\}$ which reliably distinguishes between 
the $\rho_i$ from $\ccal$ in the sense that 
\be
p_s={1\over |\ccal|}\sum_{i\in \ccal} \tr \rho_i Y_i \geq 1- 2(\ep +\sqrt {8 \ep} ) 
- 4 \times 2^{-n\delta}
\ee
\end{lemma}

In our case the role of $\rho_i$ will be played by output Bob 
states $\sigma_B^\alpha$. However we cannot use the lemma directly, 
as in some case we do not have information about those state for each 
particular $\alpha$.  Rather, we have information about some average. 

Our goal is to evaluate probability of success $p_s$ for distinguishing randomly chosen 
$N$ states $\sigma_\alpha^B$. Instead, we will first 
evaluate average of the following quantity
\be
\tilde p_s={1\over N} \sum_\alpha \tr \sigalpb Y_\alpha
\ee
where $Y_\alpha$ is some suitably chosen POVM. Since $\sigalpb$ are not 
normalized, this does not have interpretation of probability of success. 
However we will now show, that $p_s$ is close to $\tilde p_s$. 
\begin{lemma}
Let $\ep>0$. Then for $n$ large enough, 
\be
p_s={1\over N} \sum_\alpha \tr(\sab Y_\alpha) \geq 
{1\over N} \sum_\alpha \tr(\sigalpb Y_\alpha) - \ep
\ee
\end{lemma}
{\bf Proof.} Using inequality $|\tr (A B )|\leq ||A || ||B||_{op}$ 
where $||\cdot||_{op}$ is operator norm, we obtain
\be
\tr [(\sab -\sigalpb) Y_\alpha] \leq  ||\sab-\sigalpb|| \times ||Y_\alpha||_{op}
\ee
Since $Y_\alpha$ are elements of POVM, we have $||Y_\alpha||_{op}\leq 1$. 
Thus we get 
\be
{1\over N} \sum_\alpha \tr(\sab Y_\alpha) \geq 
{1\over N} \sum_\alpha \tr(\sigalpb Y_\alpha) - 
{1\over N} \sum_\alpha||\sab-\sigalpb||. 
\ee
From lemma \ref{lem:sig-sig} we know that the last term is arbitrary small 
for $n$ large enough, hence we obtain the required estimate. \blacksquare

Thus it is enough to estimate the quantity $\tilde p_s$. 
Below we will show that on average the quantity $\tilde p_e =1- \tilde p_s$ 
is exponentially small, provided $N=2^{nR}$, with $R<\ic$. 

\begin{proposition}
Fix arbitrary $\ep>0$  and $n$ large enough. Consider 
code $\ccal$ consisting of $N=2^{nR}$ codewords $|\alpha\>$ where $R<\ic$.  
Then there exists POVM 
$\{Y_\alpha\}$ such that 
\be
\lav {1\over N} \sum_\alpha \tr(\sigalpb Y_\alpha)\rav_\ccal \leq \ep.
\ee
\end{proposition}

{\bf Proof.} Let us recall important for us properties of states $\sigalpb$.
We have 
\ben
&&\rk \sigalpb=\rk\sigalpe=d_E\leq  2^{-n(S_E-\delta)},
\label{eq:rank-sigb}\\
&&\Pi_B \sigalpb \Pi_B=\sigalpb,
\label{eq:pib-sigb}\\
&&\int \sigalpb\dt \alpha = \rhonb
\label{eq:av-sigb}
\een

For a  fixed $N$ element code $\ccal$, we have the correspoding  set of states 
$\sigalpb$, we choose POVM as 
\be
Y_\alpha=(\sum_{\beta\in\ccal} \Lambda_{\beta})^{-1/2} \Lambda_\alpha 
(\sum_{\beta\in\ccal} \Lambda_{\beta})^{-1/2}
\ee
with 
\be
\Lambda_\alpha=\Pi^\alpha_B \Pi_B \Pi^\alpha_B
\ee
where $\Pi^\alpha_B$ is a projector onto support of $\sigalpb$. 
We then know that $\tr \Pi^\alpha_B\leq 2^{-n(S_E-\delta)}$. 
We will use operator inequality proven by Nagaoka and Hayashi
\cite{HayashiN2002-HSW}
\be
I-(S+T)^{-1/2} S (S+T)^{-1/2} \leq 2(1-S) + 4 T
\ee
valid for any operators satisfying $0\leq S\leq I$ and $T\geq 0$. 
Using it we can evaluate average of $\tilde p_e $ over codes as follows 
\be
\<\tilde p_e \>_\ccal=1- \lav {1\over N } \sum_\alpha \tr \sigalpb Y_\alpha \rav_\alpha
\leq 2(1 - \<\tr (\sigalpb \Lambda_\alpha)\>_\alpha) + 4 \sum_{\beta\not = \alpha}\<\tr 
(\sigalpb \Lambda_{\beta})\>_{\alpha,\beta}
\ee
Using (\ref{eq:pib-sigb}) we get $\tr (\sigalpb \Lambda_\alpha)=\tr \sigalpb$ 
so that $\<\tr (\sigalpb \Lambda_\alpha\>_\alpha=\tr \rhonb$.
Since $\beta's$ are drawn independently, 
\be
\sum_{\beta\not = \alpha}\<\tr  \sigalpb \Lambda_{\beta}\>_{\alpha,\beta}
=(N-1) \tr (\rhonb \<\Pi^\alpha_B\>_\alpha) \leq (N-1) ||\rhonb||_{op} \<\tr \Pi^\alpha_B\>_\alpha
\leq (N-1) 2^{n(S_E+\delta)} (1+\ep) 2^{-n(S_B-\delta)}
\ee
where we have used properties of state $\rhonb$ form section \ref{subsec:rhonx}.
we see that if $R<S_B-S_E$ then $\tilde p_e$ goes down exponentially. This ends 
the proof of proposition \blacksquare

In this way we have proven second ingredient of the quantum coding theorem. 
It is instructive to see what properties of code we have used in this 
section. We have based solely on the fact that the code ensemble 
gives rise to density matrix $\rhotypan$, and that the actual state $\psi_{ABE}$ 
is close to projected state $\tilde\psi_{ABE}$.
We have shown in section \ref{sec:psi-actual} for normalized codes, the latter 
property follows from the former one. 
Thus for normalized codes, distignuishability by Bob 
is guaranteed by the fact that we draw codes from ensemble of 
typical version of course density matrix. 
For unnormalized codes, the condition "actual close to projected" 
needs to be checked, but once it is satisfied for a given code, 
we obtain distinguishability again for free.

\section{Conclusions}
In this paper we have considered quantum coding theorem 
with random quantum codes. We have managed to divide the problem 
into two subproblems: checking that codewords are distinguishable by receiver (Bob)
and indistinguishable by the one who controls environment (Eve). 
For normalized codes, we have shown that distinguishability by Bob 
is  matched, when we draw code  form whatever ensemble 
of typical version of source density matrix. 
In contrast, we have not exhibited general conditions, which ensure 
privacy (indistiguishability by Eve). We have checked them 
for two types of codes which we have considered. 
We hope that this paper will also provide insight 
into problem of coding in more general type of channels.

\subsection*{Acknowledgements}
MH is supported by Polish Ministry of Scientific Research and Information Technology under the (solicited) grant no. PBZ-MIN-008/P03/2003 and  EC IP SCALA.
SL acknowledges the support by NEC and W. M. Keck foundation. 
AW is supported by the U.K. EPSRC (project "QIP IRC" and the
Advanced Research Fellowship), by a Royal Society Wolfson Merit Award,
and the EC, IP "QAP".

\section{Appendix}
\subsection{Supports}
\begin{lemma}
\label{lem:supp}
For any states $\sigma,\rho$ if 
$\supp(\sigma)\subset\supp(\rho)$ then for any CP map also 
$\supp(\Lambda(\sigma))\subset \supp(\Lambda(\rho))$. 
\end{lemma}
{\bf Proof.} We leave this ass an exercise for the reader. 

\subsection{Norms and fidelity}
The following lemma that relates trace norm with Hilbert Schmidt norm
\begin{lemma}
\label{lem:norms}
For any hermitian operator $X$ we have 
\be
||X||^2 \leq d ||X||_{HS}^2
\ee
where $d$ is dimension of the support of operator $X$ (the subspace on which 
$X$ has nonzero eigenvalues), and $|| \cdot ||_{HS}$ is Hilbert-Schmidt norm,
given by 
\be
||X||_{HS}= \sqrt {\tr X^2}
\ee
for any Hermitian operator $X$, and $||\cdot||$ is trace norm.
\end{lemma}
{\bf Proof.} It it implied by convexity of function $x^2$, where 
one takes probabilities $1/d$. \blacksquare

The fidelity given by 
\be
F(\rho,\sigma)=\tr \sqrt{\sqrt \rho \sigma \sqrt \rho}
\ee
is related to trace norm as follows \cite{Fuchs-Graaf} 
\be
1-F \leq  {1\over 2} ||\rho-\sigma||\leq \sqrt (1- F(\rho,\sigma))^2.
\label{eq:fuchs}
\ee

We will also need the following lemma
\begin{lemma}
\label{lem:psi-psit}
For any vector $\psi'$ and $\psit=\Pi\psi'$ where $\Pi$ is a projector, where 
$|\<\psit|\psi'\>|^2\geq 1-\ep$,  $\<\psi'|\psi'\>\leq 1+\ep$ and  $0<\ep<1$ we have 
\be
||\,|\psit\>\<\psit| - |\psi\>\<\psi|\, ||\leq 6\sqrt {\ep}
\ee
where $\psi$ denotes normalized version of $\psi'$, i.e. $\psi=\psi'/\sqrt{\<\psi'|\psi'\>}$
\end{lemma}

{\bf Proof.} For any vector $|v\>$ denote $P_v=|v\>\<v|$. We then have 
\be
||P_\psit-P_\psi|| \leq ||P_\psit - {P_\psit\over \tr P_\psit}|| + 
||{P_\psit\over \tr P_\psit}-P_\psi||\leq |1-\<\psit|\psit\>| 
+ 2\sqrt{1- {|\<\psit|\psi\>|^2 \over \<\psit|\psit\>}}
\ee
where we have used inequality (\ref{eq:fuchs}). Subsequently, we get 
\be
||P_\psit-P_\psi|| \leq |1-\<\psit|\psit\>| 
+ 2\sqrt{1- {|\<\psit|\psi'\>|^2 \over \<\psi'|\psi'\>}}
\ee
Now, form assumptions of the lemma, it follows that $(1-\ep)/(1+\ep)\leq  \<\psit|\psit\>
\leq 1+\ep$. This together with the assumptions gives 
\be
||P_\psit-P_\psi||\leq 2\ep + 2 \sqrt{1- {1-\ep \over (1+\ep)^2}}\leq 6\sqrt \ep
\ee
This ends the proof of the lemma. \blacksquare

\subsection{Averaging over codewords}
In this section we will present useful formulas for averaging over codewords. 

From \cite{Lloyd-cap} we get the following rules for Lloyd codes.
\ben
&\displaystyle\int \<\alpha | X | \alpha\> \<\alpha|Y |\alpha\> \dt \alpha= &
\sum_{ij} p_ip_j \<i|X|i\>\<j|Y|j\> +
\sum_{ij} p_ip_j \<i|X|j\>\<j|Y|i\> 
- \sum_{i} p_i^2\<i|X|i\>\<i|Y|i\> =  \nonumber \\  
&&=\tr (X \rhotypan) \tr (Y \rhotypan) + \tr (X\rhotypan Y\rhotypan )
-\sum_{i} p_i^2\<i|X|i\>\<i|Y|i\>
\label{eq:aaaa-s}
\een
for any operators $X,Y$.  
For uniform source-distorted codes, we use properties of 
projector onto symmetric subspace and get 
\ben
\displaystyle\int \<\alpha | X | \alpha\> \<\alpha|Y |\alpha\> \dt \alpha
={d_A\over d_A+1 }\bigl[\tr (X \rhotypan) \tr (Y \rhotypan) + \tr (X\rhota Y\rhotypan )\bigr]
\label{eq:aaaa-a}
\een
for any operators $X,Y$.  
For both codes we have 
\be
\int |\alpha\>\<\alpha| \dt \alpha= \rhotypan
\label{eq:aa}
\ee

\bibliographystyle{apsrev}
\bibliography{refmich,refpostmich}


\end{document}